\documentclass[12pt]{article}
\usepackage{epsfig}
\newcommand{\R}{I\!\!R}
\newcommand{\C}{l\!\!\!C}

\newcommand{\noi}{\noindent}

\newcommand{\la}{\langle}
\newcommand{\ra}{\rangle}

\newcommand{\rhotilde}{\tilde{\rho}}
\newcommand{\rhobar}{\bar{\rho}}
\newcommand{\rhoeq}{\rho_{eq}}
\renewcommand{\hbar}{\bar{h}}
\newcommand{\htilde}{\tilde{h}}

\newcommand{\stilde}{\tilde{s}}
\newcommand{\lambdadot}{\dot{\lambda}}
\newcommand{\lambdahat}{\hat{\lambda}}
\newcommand{\ahat}{\hat{a}}

\renewcommand{\d}{\partial}
\newcommand{\half}{\frac{1}{2}}

\renewcommand{\L}{{\cal L}}
\renewcommand{\H}{{\cal H}}

\begin{document}

\title{{\bf An optimization principle for deriving nonequilibrium
statistical models of Hamiltonian dynamics } }
\author{ Bruce Turkington 
  \\  
  \small Department of Mathematics and Statistics  \\
  \small University of Massachusetts Amherst    }

\date{}
\maketitle
\normalsize

\begin{abstract}
A general method for deriving closed reduced models of Hamiltonian dynamical systems is
developed using techniques from optimization and statistical estimation.   
As in the standard projection operator methodology of statistical mechanics, 
a set of resolved variables is selected to capture the slow, macroscopic behavior of the
system, and the family of quasi-equilibrium probability densities  on phase space
corresponding to these resolved variables is employed as a statistical model.  
The macroscopic dynamics of the mean resolved variables is determined by optimizing
over paths of these probability densities.  Specifically, a cost function is introduced
that quantifies the lack-of-fit of a feasible path to the underlying microscopic dynamics;
it is an ensemble-averaged, squared-norm of the residual that results from submitting
a path of trial densities to the Liouville equation.   The evolution of the macrostate
is estimated by minimizing the time integral of the cost function over such paths.  
Thus, the defining principle for the reduced model takes the form of Hamilton's principle 
in mechanics, in which the Lagrangian is the cost function and the configuration variables
are the parameters of the statistical model.    
The value function for this optimization, which plays the role of the action integral,  satisfies the
 associated Hamilton-Jacobi equation,  and it determines the optimal relation between
 the statistical parameters and the irreversible fluxes of the resolved variables, thereby
 closing the reduced dynamics.    The resulting equations for the macroscopic variables have
 the generic form of governing equations for nonequilibrium thermodynamics, and 
 they furnish a rational extension of the classical equations 
 of linear irreversible thermodynamics beyond the near-equilibrium regime. 
 In particular, the value function is a thermodynamic potential that
 extends the classical dissipation function and supplies the nonlinear relation between
 thermodynamics forces and fluxes.

\end{abstract}

\noi {\it Key Words and Phrases: 
model reduction, statistical closure, dynamical optimization, nonequilibrium thermodynamics.}

\section{Introduction}

A main goal of nonequilibrium statistical mechanics is the derivation
of effective equations for macroscopic behavior from the equations of motion
that govern an underlying microscopic dynamics \cite{Balescu,Dougherty,Katz,Keizer,Penrose,
Ramshaw,Zwanzig}.     Unlike equilibrium statistical
mechanics, which rests on the firm foundation of conserved variables and Gibbsian ensembles, 
this endeavor is necessarily imperfect and approximate.  First, it relies on the choice of 
relevant macroscopic variables, which is not universal.  Even though in traditional settings
such as hydrodynamics, the local equilibrium hypothesis can be invoked to justify
the use of locally conserved quantities, in many situations it may be advantageous 
to consider reduced or extended sets of relevant variables \cite{JC-VL}. 
Second,  the selection of a macroscopic description 
depends on the existence of a separation of time scales between the relevant
variables that evolve on a slow time scale and the remaining microscopic variables that
are expected to equilibrate on fast time scales.    But, in some problems for which a statistical
mechanical description is sought, such as fluid turbulence,   a wide separation between
fast and slow variables may not be present \cite{McComb}.  Third, when equations of motion for
the relevant variables are derived by standard projection operator methods, they contain
finite-memory effects and non-Markovian stochastic forcing terms, which are clumsy to
handle until further approximations or limits are adopted to bring them to 
more tractable forms \cite{CHK1,CHK2}.   As a consequence, most statistical treatments of nonequilibrium
phenomena are restricted to idealized physical models having special microscopic
dynamics.    Often a key simplification is to replace a deterministic microscopic dynamics
by a stochastic process at the outset of the analysis, constructing
the process to be compatible with equilibrium fluctuations \cite{GKS}.                   

In the present paper we take a different approach, in that we seek to characterize 
the governing equations for a given set of relevant,
or resolved, variables by applying a statistical model reduction procedure to the 
underlying deterministic dynamics itself, which we take to be a general Hamiltonian  system
having finitely-many degrees of freedom.       
The central idea in our approach is a dynamical optimization principle over paths
of macrostates, which may be outlined as follows.  (i)  Relative to a given set of 
resolved variables, a parametric statistical model of the system is imposed using
quasi-equilibrium probability
densities with parameters conjugate to the expectations of the resolved variables. 
(ii)  A macroscopic evolution is identified with a path in the parameter space of
these trial probability densities, and
the lack-of-fit of such a path to the underlying Hamiltonian dynamics is quantified
by a certain cost functional, which is a time-integrated, weighted, squared norm 
of the residual of those densities with respect to the Liouville equation.
(iii)   The estimated, or predicted,  evolution of the macrostate is determined by that
path which minimizes the lack-of-fit cost functional.       
In this way we derive an
irreversible, dissipative dynamics that is closed in the macrovariables and is optimally compatible
with the underlying conservative, reversible microdynamics  in a quantified, statistical sense.       

A particularly desirable feature of the closed reduced equations derived via our approach
is that they have the structure of the general equations for nonequilibrium
thermodynamics proposed by contemporary researchers 
\cite{Morrison,GO,OG,Ottinger-book,BE,Berdichevsky}.       
In particular, our model reduction
strategy may be viewed as a dynamical variational principle that justifies the so-called 
GENERIC (General Equations
of NonEquilibrium Reversible Irreversible Coupling) format for out-of-equilibrium dynamics 
\cite{Ottinger-book}.    
In the notation used in the body of our paper, these equations have the form
\begin{equation} \label{generic-intro}  
\frac {d a_i} {dt} \, = \, \sum_{j} \Omega_{ij} \frac{\d h}{\d a_j} \; -  \;  \frac{\d v }{\d \lambda^i} \, , 
\end{equation}  
in which $a=(a_1, \ldots , a_m)$ denotes the macrostate vector, which is the expectation
of the given vector of resolved variables $A = (A_1, \ldots , A_m)$ defined on the phase space
of the Hamiltonian dynamics.   The vector 
$\lambda= (\lambda^1, \ldots, \lambda^m)$
consists of parameters conjugate to $a$, given by $\lambda^i = - \d s / \d a_i$, where
$s=s(a)$ is the entropy of the macrostate $a$.    
 The first term in (\ref{generic-intro}) is a generalized Hamiltonian vector
field having an energy function $h=h(s, a)$ and a Poisson matrix $\Omega = \Omega(a)$, 
which is anti-symmetric;  the partial derivatives $\d h / \d a_i $ are at constant entropy $s=s(a)$.   
This term governs the reversible part of the reduced dynamics.    The second term 
in (\ref{generic-intro}) is a gradient vector field with respect to the parameter 
vector $\lambda$ having the potential function $v=v(\lambda)$.   
 This term governs the irreversible part of the reduced dynamics.    In our variational formulation
the thermodynamic potential $v(\lambda)$ is the value function for the optimization
principle; that is, $v(\lambda)$ equals the optimal value of the lack-of-fit cost
functional  in the minimization over paths emanating from the macrostate with
parameter vector $\lambda$ and tending to equilibrium in infinite time.  
The value function therefore solves the time-independent Hamilton-Jacobi equation
associated with optimization principle.       
The functions $h(s,a)$ and $\Omega(a)$ in the reversible term are entirely dictated
by the form of the trial probability densities defining the statistical model.
By contrast,
the value function $v(\lambda)$ depends on the weights introduced into the lack-of-fit cost function,
and these weights determine the adjustable parameters of the ensuing closure theory.

Under the near-equilibrium approximation, the closed reduced equations    (\ref{generic-intro})
take the classical form of linear irreversible thermodynamics \cite{dGM}.    Namely, a linear system of
differential equations relates the fluxes $d a_i/ dt$ to the affinities (or thermodynamic forces)
$- \lambda^j$;    namely,  
\begin{equation}   \label{LIT-intro}
\frac {d a_i} {dt} \, = \, \sum_{j} \left[  \, \beta^{-1} \Omega_{ij} - M_{ij} \,   \right] \, \lambda^j  \, , 
 \hspace{2cm}   a_i = \sum_j C_{ij} \lambda^j \, . 
\end{equation}  
The coefficients in these equations are evaluated at equilibrium, $a=0$;   
$\beta$ is the inverse temperature, and 
the matrix $(C_{ij})$ is the inverse of the matrix $ - (  \d^2 s / \d a_i \d a_j  )$. 
In this formula $M$ is the symmetric and
positive-definite matrix that defines the near-equilibrium value function, which is the quadratic form
\[
v(\lambda) \, = \,   \half \sum_{i,j} M_{ij}  \lambda^i  \lambda^j  \, .   
\]
The physical meaning of the value function becomes transparent from these identities, 
which imply that  $ d s / dt = 2 v(\lambda) $.  That is, the optimal value of the cost functional coincides
with half the entropy production along the best-fit path.  

Beyond the regime in which these near equilibrium
approximations hold, the closed reduced equations (\ref{generic-intro}) remain valid as the
nonlinear equations that determine the best fit of the statistical model to the underlying dynamics.   
The utility of these equations depends on the choice of the resolved variables and the
separation of time scales between the resolved and unresolved variables in the statistical
model, not on the closeness of the modeled macrostates to equilibrium.  
In the far-from-equilibrium regime the value function $v(\lambda)$ furnishes a natural
extension of the classical dissipation function that is justified by a statistical mechanical derivation.
Accordingly, our optimization-based derivation of the GENERIC theory offers both  a new
interpretation of the potential for the irreversible part of the governing equations (\ref{generic-intro}), and
a systematic way to calculate nonlinear corrections to the classical linear equations  (\ref{LIT-intro}).

The paper is organized as follows.   Section 2 presents the background on the general
closure problem and introduces the family of trial probability densities that
constitute the statistical model.     In Section 3 the stationary version of the defining
optimization principle is formulated and the general nonequilibrium equations for the
 best-fit reduced model are derived.    Properties of these equations are discussed in
Section 4, as well as the near-equilibrium linearization of the
reduced model.    In Section 5 the theory is extended to include an initial period of 
development during which the entropy production increases from zero.    This plateau
effect is encountered when the specified initial statistical state is itself a quasi-equilibrium ensemble.
The  value function then becomes time-dependent, $v=v(\lambda,t)$, and solves a time-dependent Hamilton-Jacobi equation.
Section 6 discusses the conceptual framework 
of the best-fit approach to closure and points to some specific test cases and applications.   

\section{Statistical model and trial densities }

The underlying microscopic dynamics is assumed to be a Hamiltonian system.
In canonical form the equations of motion are 
\begin{equation}  \label{hamilton-eqn}
\frac{dz}{dt} \,=\, J \nabla_z H(z) \;\;\;\; \mbox{ with } \;\;\;\; J
\,=\, \left( \begin{array}{ccc} O & I \\ -I & O
  \end{array}  \right) \, ,  
\end{equation}
where $z=(q,p)$ denotes a generic point in the phase space $\Gamma_n =
\R^{2n}$, $n$ being the number of degrees of
freedom of the system \cite{Arnold,Lanczos}.  Our theoretical development permits general systems
of this kind, with time-independent, smooth Hamiltonians $H$ on $\Gamma_n$.
In fact,  our development only makes use of the general structure of the underlying
microscopic dynamics --- essentially, the conservation of energy and phase volume. 
Consequently it also applies without fundamental modifications to dynamical 
systems in a noncanonical Hamiltonian form.      It is many such noncanonical systems, however,
there are conserved quantities other than $H$ which play an important role 
in conditioning the statistical behavior of the dynamics.    
A version of the theory appropriate to infinite-dimensional dynamics,
such as occur in continuum models, could also be formulated.   But it would be
necessary to carry out a limiting analysis of discretized models in order to justify
the statistical description.    For the sake of definiteness, therefore, we 
restrict our discussion to finite-dimensional, canonical systems (\ref{hamilton-eqn}).

We are interested in making predictive estimates of some relevant macroscopic
dynamical variables rather than
following the details of the microscopic dynamics (\ref{hamilton-eqn})
itself.  We therefore seek a statistical closure in terms of some resolved dynamical variables. 
In keeping with the general character of our approach, we allow any finite
number of arbitrary resolved variables  $A_k$, $k=1, \ldots , m$,  assuming that each is a
smooth real-valued function on $\Gamma_n$ and the set $A_1,
\ldots , A_m$ is linearly independent.   We assemble them into the 
resolved vector $A=(A_1, \ldots ,A_m)$.    Typically,  $m \ll n$.

The evolution of any dynamical variable, $F$, resolved or not,
is determined by the equation
\begin{equation}  \label{poisson-eqn} 
\frac{dF}{dt} \,=\, \{ F , H \} \, ,
\end{equation}
where $\{F,H\} = (\nabla F)^* J \nabla H$ is the Poisson bracket
associated with the canonical Hamiltonian structure.  Indeed, the
statement that (\ref{poisson-eqn}) holds for all smooth functions $F$
on $\Gamma_n$ is equivalent to the Hamiltonian dynamics
(\ref{hamilton-eqn}).  Fundamentally, the problem of closure in terms
of the resolved variables $A_1, \ldots, A_m$ arises from the fact
that, except under very special circumstances, the derived
variables $ \{A_1,H\}, \ldots,  \{A_m,H\}$ are
not expressible as functions of the resolved variables $A_1, \ldots, A_m$.  
This generic fact makes a statistical description of the unresolved variables necessary.  
The foundation for such a statistical description is provided by the Liouville
equation, which governs  the propagation of probability under the phase flow. 
Namely, for probability measures on $\Gamma_n$, $p(dz,t) =
\rho(z,t)dz$, having smooth densities $\rho$ with respect to the
invariant, $2n$-dimensional, phase volume element  $dz=dqdp$,
\begin{equation}  \label{liouville-eqn} 
\frac{\partial \rho}{\partial t} \,+\, L \rho \,=\,0 
\;\;\;\; \mbox{ in } \Gamma_n \times \R \, ,    
\end{equation}
in which  $L= \{ \cdot , H \}$ denotes the Liouville operator.    
Given a density $\rho(z,t_0)$ at an initial time $t_0$,
(\ref{liouville-eqn}) completely determines the density $\rho(z,t)$ at
any later time $t$, which is denoted formally by $\rho(\cdot, t)
= e^{- (t-t_0) L} \, \rho(\cdot, t_0)$.  We denote the expectation of
any dynamical variable $F$ with respect to $\rho(t)$ by
\[
\la F | \rho(t) \ra \,=\, \int_{\Gamma_n} F(z) \, \rho(z,t) \, dz \, .   
\]
[This bracket notation is used for expectation throughout the paper.]
The Liouville equation is equivalent to the statement that 
\begin{equation}  \label{moment-eqn}
\frac{d}{dt}  \la F | \rho(t) \ra \,=\, \la L F | \rho(t) \ra \,   
\end{equation}
for every dynamical variable $F$.    
In particular, the evolution of the mean of the
resolved vector, $a(t)= \la A | \rho(t) \ra$, is determined by the exact
solution of (\ref{liouville-eqn}).  But the exact density
$\rho(z,t)$ evolving under the Liouville equation is highly intricate;
it contains vastly more information than the resolved $m$-vector
$a(t)$; and, it is excessively expensive to compute numerically,
since it requires the integration of
the microscopic trajectories starting from each sample point in
an initial ensemble that approximates the 
given initial density.  

For this reason we seek a statistical closure in terms of 
an analytically tractable family of trial probability densities having
the same level of complexity as the resolved variables themselves.  
In the nomenclature of statistical inference, we impose a parametric 
statistical model for which the resolved vector $A$ is a minimal sufficient statistic
\cite{CB,Kullback}.     
We denote this family of probability densities by  $\rhotilde(z;\lambda)$, using a
parameter vector $\lambda = (\lambda^1, \ldots , \lambda^m) \in \R^m$.      
The family is assumed to be regular, meaning that 
the densities depend smoothly on $\lambda$, the
score variables are defined by 
\begin{equation}    \label{score}
U(\lambda) = \frac{\d \log \rhotilde(\lambda)}{\d \lambda}, 
\end{equation}
and the Fisher information matrix defined by 
\begin{equation}  \label{fisher}
C(\lambda)  =  \la U(\lambda)   U(\lambda)^*  | \rhotilde(\lambda)   \ra   
\end{equation}
is nonsingular.     [Here and throughout the paper we use the notation
$M^*$ to denote the transpose of any matrix $M$; in particular, a
column vector $V$ is taken to its dual row vector $V^*$.    Also, we use
the shorthand notation $\d / \d \lambda = (\d / \d \lambda^1, \ldots , \d / \d \lambda^m)$.]
The assumption that $A$ is a minimal sufficient statistic for the family $\rhotilde(\lambda)$
implies that there is a one-to-one correspondence between the expected
resolved vector $a=\la A | \rhotilde(\lambda) \ra$ and the parameter vector $\lambda$.
Accordingly, each macrostate can be identified with a unique parameter vector $\lambda$, which
ranges over the configuration space $R^m$, or perhaps a subset of it. 
We assume that the parameterization
in terms of $\lambda$ is arranged so that the origin, $\lambda =0$, is an equilibrium
density, meaning that $L \rhotilde(0) = 0. $     

Our interest centers on modeling the relaxation of 
 the  macrostate, $a(t) = \la A | \rhotilde(\lambda(t)) \ra$,  
 from a specified nonequilibrium macrostate $a^0$
at time $t=0$ towards equilibrium.  Such a relaxation corresponds to a path $\lambda(t)$ 
in parameter space for which  $\lambda(0)=\lambda_0 \neq 0$ and 
$\lambda(t) \rightarrow 0$ as $t \rightarrow +\infty$.  

Even though we formulate the best-fit closure concept in such a general setting, we 
concentrate our attention in this paper on the model that uses the family of 
 quasi-equilibrium, or quasi-canonical, densities
\cite{Jaynes1,Jaynes2,LVR,Zubarev}, 
\begin{equation}  \label{canonical-density}
\rhobar(z;\beta, \lambda) \,=\, 
  \exp [ -\beta H + \lambda^*A - \psi(\beta,\lambda) ]   \, ,
\end{equation}
in which 
\begin{equation}   \label{psi}
  \psi(\beta, \lambda) \,=\, 
         \log \int_{\Gamma_n}  \exp ( - \beta H + \lambda^*A) \, dz \, .
\end{equation} 
[Here and throughout, $\lambda^*A = \sum_{i} \lambda^i A_i$.]   These densities
depend on the inverse temperature $\beta$ as well as the parameters 
$\lambda^1, \ldots, \lambda^m$ associated with the nonconserved resolved variables.   
Some growth conditions on $H$ and $A$ at infinity 
may be required to ensure that the normalization (\ref{psi})  exists and is finite; 
for simplicity in our exposition, we assume that these densities exist for all $\beta >0$
and $\lambda \in \R^m$.     
The associated entropy function, $s(u,a)$, 
of the mean energy, $u$, and the mean resolved vector $a$, is determined by the maximum entropy principle,
\begin{equation}   \label{max-ent}
s(u,a) \, = \, \max_{\rho} \; -  \la \, \log \rho \, | \, \rho \, \ra   \;\;\;\;\;
   \mbox{ subject to } \;\; \la  H  |  \rho \, \ra \, = u , \;\;  \la  A |  \rho \ra  \, =a , 
    \;\;  \la 1 |  \rho  \ra \, = 1 \, . 
\end{equation}
The maximizer in (\ref{max-ent}) is the quasi-canonical density (\ref{canonical-density}),
and the parameter $\beta$ and parameter vector $-\lambda$ are Lagrange multipliers for the 
constraints on mean energy and mean resolved vector, respectively.       
The entropy function $s(u,a)$ is the negative of the convex conjugate function to $\psi(\beta,\lambda)$ 
defined in (\ref{psi}); that is, $-s(u,a)$ is the Legendre transform of $\psi(\beta, \lambda)$,
and hence all the following relations hold:
\[
-  s(u,a) = - \beta u + \lambda^*a - \psi(\beta,\lambda)  \, ,  \;\;\;\;
 - \frac{\d \psi}{\d \beta} = u \, , \;\;  \frac{\d \psi}{\d \lambda} = a \, , \;\;\;\;
\frac{\d s}{\d u} = \beta \, , \;\;   - \frac{\d s}{\d a} = \lambda \, . 
\]

There are two ways to handle the dependence of the family (\ref{canonical-density})
on $\beta$, or equivalently, on the mean energy $u = \la H | \rhobar \ra$, and
this choice leads to two versions of the best-fit closure theory.       

In one version we fix $\beta>0$ and set $\rhotilde(\lambda) = \rhobar(\beta,\lambda)$.   
The equilibrium density $\rhotilde(0)$ is the canonical Gibbs ensemble on $\Gamma_n$,
\begin{equation}  \label{gibbs}
\rho_{eq} (z) \, = \, Z(\beta)^{-1}\exp ( - \beta H (z) \, ) \, ,   \;\;\;\;\;\;\;\; 
\mbox{ with }  \;\; Z (\beta) = \int_{\Gamma_n} \exp ( - \beta H (z) \, ) \, dz \, .
\end{equation} 
We let $\la F \ra_{eq}$ denote expectation of any dynamical variable $F$ with 
respect to  $\rho_{eq}$.    The canonical statistical model
with fixed inverse temperature therefore consists of the densities 
\begin{equation}   \label{qe-density}
\rhotilde(z;\lambda) \,=\, \exp ( \lambda^*A - \phi(\lambda)) \, \rhoeq(z) \, ,
\;\;\;\;\; \mbox{ with } \;\;  \phi(\lambda) \,=\, 
         \log \, \la \,  \exp (\lambda^*A) \, \ra_{eq} \, .
\end{equation} 
Under this choice of trial densities the mean energy $\la H | \rhotilde(\lambda) \ra$
is allowed to vary along a path $\lambda=\lambda(t)$.   This version of the model
is appropriate to a physical system in which an energy  reservoir maintains
a constant system temperature.

A standard motivation for using   the family of densities (\ref{qe-density}) 
is that each member of the family minimizes relative entropy subject to
the mean value of the resolved vector; that is, $\rhotilde$ solves
\[
- \stilde(a) \, = \,  \min_{\rho}  \; \la \, \log \frac{ \rho }{\rhoeq} \, | \, \rho  \, \ra
\;\;\;\;\;\; \mbox{ subject to } \; \la  A | \rho \ra \,
= \, a \, , \;\;  \la 1 | \rho \ra \, \,=\,1.
\]
From the perspective of information theory \cite{CT,Kullback},
$\rhotilde(z;\lambda)$ is the least informative probability density
relative to the equilibrium density $\rhoeq$ that is
compatible with the macrostate vector
$ a \,=\, \la A | \rhotilde \ra $.   
The relative entropy, $- \stilde(a)$, as a function of the mean resolved vector, $a$,
is the Legendre transform of $\phi(\lambda)$, and 
the one-to-one
correspondence between $\lambda$ and $a$ is a convex duality given by
\begin{equation}   \label{qe-duality}
a \,=\, \frac{\partial \phi}{\partial \lambda} \, ,  \;\;\;\;\;\;
\lambda \,=\, - \frac{\partial \stilde}{\partial a} \, .   
\end{equation}

In statistical inference, (\ref{qe-density}) is called an exponential family relative to $\rhoeq$ with
natural parameter $\lambda$ \cite{CB}.   The score vector,  $U(\lambda) \, = \, A - a  $,
is the resolved vector centered around its mean, $a=\la A | \rhotilde(\lambda) \ra$.   
The Fisher information is $C(\lambda) = \la \,  (A-a) (A-a)^* \, | \, \rhotilde (\lambda) \, \ra $,
the covariance matrix for the resolved variables.     

In the alternative version we impose the conservation of mean energy 
$E = \la H | \rhotilde (\lambda(t)) \ra$ exactly along paths $\lambda(t)$.
This constraint is achieved by allowing $\beta = \beta(\lambda)$ to vary with
$\lambda$, and setting $\rhotilde(\lambda) = \rhobar(\beta(\lambda), \lambda)$.   
From a physical perspective this version is appropriate whenever the model system is isolated.
Moreover, it leads to a closed reduced dynamics 
having precisely the form of GENERIC thermodynamics \cite{Ottinger-book}.
For these reasons, we also develop this version of the best-fit closure theory  
in the present paper, even though the version with fixed temperature is technically
simpler.     

The function $\beta(\lambda)$ is determined by the requirement that,
for all admissible parameter vectors $\lambda$, 
\begin{equation}  \label{conserve-energy}
\la \, H \, | \, \rhobar(\beta,\lambda) \, \ra =\, -  \frac{\d \psi}{\d \beta}(\beta,\lambda) \, = \, E ,  
\end{equation}
 where $E$ is the equilibrium energy corresponding to $\lambda=0$.   
The unique solvability of this equation for $\beta = \beta(\lambda)$ is ensured by the
strict convexity of $\psi$.    
Differentiating (\ref{conserve-energy}) with respect to $\lambda$ yields
\[
0 = \frac{\d }{\d \lambda} \la H | \rhotilde(\lambda) \ra \, = \,  
   \la (H - E) U(\lambda) | \rhotilde(\lambda) \ra   \, 
\]
where the score vector in this energy-conserving model is
\begin{equation}   \label{score-energy}
U(\lambda) \, = \,  (A - a) \, - 
    \frac{\la (H-E) (A -a)  | \rhotilde \ra } {  \la  (H-E)^2 | \rhotilde \ra}   (H-E)   \, .   
\end{equation} 
Thus, the score variables $U_i(\lambda) $ are the centered resolved variables $A_i$ 
projected onto the subspace of $L^2(\Gamma_n, \rhotilde)$ orthogonal to
the centered Hamiltonian $H-E$.    The Fisher information matrix (\ref{fisher}) is the covariance matrix
for these projected resolved variables:
\begin{equation}   \label{fisher-energy}  
C(\lambda) \, = \,     \la \,  (A-a)(A-a)^* \, \ra \, - \, 
                \frac{\la  (H-E) (A -a) \ra  \la (H-E) (A -a)^*  \ra} { \la (H-E)^2 \ra }  \, . 
\end{equation}

Before proceeding to formulate our optimization principle over paths of macrostates,
let us first review a naive closure procedure for the quasi-equilibrium probability
densities $\rhotilde(\lambda)$ that produces the reversible part of
the closed reduced dynamics, but entirely suppresses the irreversible part \cite{GK}.
Motivated by the moment equations (\ref{moment-eqn}), which hold for any
dynamical variable $F$ and for the exact Liouville solution $\rho(\cdot, t)$, 
one may impose the $A$-moments of the Liouville equation on the trial densities
$\rhotilde(\lambda(t))$.  Then the parameter path $\lambda(t) \in \R^m$
is required to satisfy the $m$ differential equations
\[
0 \, = \, \la \, A \, | \, \frac{\d \rhotilde}{\d t} + L \rhotilde  \,  \ra   \, = \, 
     \frac{d}{dt} \la \, A \, | \, \rhotilde \, \ra \, - \, \la  \, LA \, | \, \rhotilde \, \ra    \, .   
\]
This simple moment closure with quasi-equilibrium densities is memoryless, and
consequently it is entropy conserving.   The entropy production calculation is:
\begin{eqnarray*}
\frac{d s } {d t } \, = \,  - \lambda^* \frac{d a} {d t }  
    &=& - \la \, L ( \lambda^*A) \, | \,  \rhotilde(\lambda)     \, \ra    \\
    &=&  - \int_{\Gamma_n} \{ \lambda^*A , H \} \, 
                     \exp[ -\beta H + \lambda^*A - \psi(\beta,\lambda) ] \, dz   \\
    &=&   - \int_{\Gamma_n} \{ \exp [ -\beta H + \lambda^*A - \psi(\beta,\lambda) ] , H \}  \, dz    \, = \, 0 \, .
\end{eqnarray*}
The last equality makes use of the general integration-by-parts identity,
\[
 \int_{\Gamma_n}  \{ F_1, F_2 \} F_3 \, dz  \, = \, 
     - \int_{\Gamma_n}  \{ F_3, F_2 \} F_1 \, dz \, ,
\]
which is also used in several calculations to follow.     

The closed reduced equations governing this adiabatic  closure  can be written as
\begin{equation}   \label{adiabatic-eqn}   
\frac{d a } {d t } \, = \, f(\lambda) \, ,   \;\;\;\;\;\;\;\;\;\;   
      \mbox{ with } \;\;\;\; \lambda = -\frac{\d s } {\d a } \, ,
\end{equation}
introducing the vector field   
\begin{eqnarray}   \label{f} 
f(\lambda) \, &=&  \,   \la \, LA \, |  \, \rhotilde(\lambda) \,   \ra   \\  \nonumber
                    &=&   \int_{\Gamma_n} \{A , H \} \,  \exp [ \, -\beta H + \lambda^*A - 
                                                                            \psi(\beta, \lambda \, ) ] \, dz   \\ \nonumber
                     &=&  - \beta^{-1} \int_{\Gamma_n} \{ A ,  \,   \exp [ \, -\beta H \, ] \, \} \,
                                  \exp [ \, \lambda^*A - \psi(\beta, \lambda) ]   \, dz     \\   \nonumber   
                     &=&  \beta^{-1}  \int_{\Gamma_n} 
                                    \{ A ,  \, \exp [ \, \lambda^*A - \psi(\beta, \lambda) ]  \, \} \,
                                         \exp [ \, -\beta H \, ]   \, dz     \\      \nonumber
                     &=& \beta^{-1}  \la \, \{  A, A^* \}  \, | \, \rhotilde(\lambda) \, \ra \, \lambda \,  .  \nonumber                                                                    
\end{eqnarray}  
In either version of the model, with fixed $\beta$ or fixed $E$, it is possible to express
$f(\lambda)$ as a generalized Hamiltonian vector field.   
In both cases the  energy representation, $u=h(s,a)$, which inverts
the entropy representation, $s= s(u,a)$, plays a central role.   
 
For the model with fixed $\beta$, we introduce the associated free energy function
\begin{equation}  \label{free-energy} 
\htilde (\beta, a) \, = \, u - \beta^{-1} s   \, .   
\end{equation} 
Technically, $\htilde$ is the negative of the convex function  conjugate to $h(s,a)$
with respect to $s$, considered as  a function of the temperature $\theta = \beta^{-1} = \d h/ \d s$. 
A  straightforward calculation using the properties of the Legendre transform yields 
\[
\frac{\d \htilde} {\d  a}   \, = \,  - \beta^{-1} \frac{\d s } {\d  a} \, = \, \beta^{-1} \lambda \, , 
\]
in which the $a$-gradient of the free energy, $\htilde$, is a constant temperature.   
Substituting this expression into  (\ref{f}), we obtain
 the adiabatic closed reduced equation (\ref{adiabatic-eqn}) in the form, 
\begin{equation}   \label{adiabatic-ham-beta} 
\frac{d a } {d t } \, = \, \Omega (a) \, \frac{ \d \htilde}{ \d a}   \, .
\end{equation}  
Here we introduce the Poisson matrix,    
\begin{equation}  \label{poisson}
\Omega  \, = \,    \la \, \{  A, A^* \}  \, | \, \rhotilde(\lambda) \, \ra \, ,
\end{equation}   
which we consider as a function of $a$, recalling that there is a smooth, invertible mapping 
between $\lambda$ and $a$.     The $m \times m$ matrix $\Omega$ is antisymmetric, 
state-dependent, and not necessarily invertible.   Consequently, the reduced dynamics 
(\ref{adiabatic-ham-beta}) is a generalized
Hamiltonian system, or Poisson system, with a possibly degenerate symplectic structure.
Nonetheless, it determines a well-defined reversible dynamics 
for any regular set of resolved variables \cite{BE,Ottinger-book}.   

For the energy-conserving model,  
entropy at fixed mean energy $E$ depends on $a$ alone:  $s= s(E,a)$.  
Differentiating the identity $E=h(s(E,a),a)$ with respect to $a$ yields 
\[
0 \,=  \, \frac{\d h  }{\d s }  \frac{\d s }{\d a }  \, + \, \frac{\d h }{\d a }  \,=\, 
   \beta^{-1} (-\lambda)  \, + \, \frac{\d h }{\d a }  \, .
\]
As in (\ref{adiabatic-ham-beta}), we thereby obtain the adiabatic closed reduced dynamics
in the form  
\begin{equation}  \label{adiabatic-ham-E}
\frac{d a } {d t } \, = \, \Omega (a) \, \frac{ \d h}{ \d a}   \, ,
\end{equation}   
in which the $a$-gradient of the mean energy, $h$, is at constant entropy.

\section{Stationary formulation }

The adiabatic closure discussed in the previous section imposes the $A$-moment
of the Liouville equation on the quasi-equilibrium probability densities $\rhotilde(\lambda)$, for
$\lambda=\lambda(t) \in \R^m$, and thereby determines $m$ ordinary differential
equations of first order for the evolution of the macrostate 
$a(t) = \la \, A \, | \,  \rhotilde(\lambda(t) \, \ra$.    This dissipationless closure is clearly deficient
as a model of the statistical behavior of the resolved vector $ A \in \R^m$, because it
is suppresses the influence of the unresolved variables on the resolved variables.   
The traditional remedy for this deficiency is to obtain closure by taking the $A$-moment
with respect to a family of probability densities on phase space having memory  
\cite{LVR,Zubarev,ZMR}.  
The customary choice is given by the so-called theory of nonequilibrium
statistical operators,  in which the densities have the form
\begin{equation}   \label{neso}
\rho_{\epsilon}(\cdot,t) \, = \,   \int_0^{\infty}  e^{- \tau L  }   \rhotilde(\, \cdot \, ;\lambda(t-\tau)) \,  
         \epsilon e^{- \epsilon \tau} \, d \tau  \, .   
\end{equation}   
Such a density $\rho_{\epsilon} (\cdot, t)$ is a functional of the path $\lambda(t)$ corresponding to
a macroscopic evolution.  It depends on the values of the resolved vector $A$ at times
prior to $t$   via the propagator $e^{-\tau L } $ for the Liouville equation in elapsed time $\tau >0$,
and includes an artificial decay rate $\epsilon >0$ for the memory.       
The nonequilibrium statistical operator (\ref{neso}) at time $t$ is thus a 
time-weighted superposition of exactly propagated solutions from quasi-equilibrium densities 
at earlier times $t-\tau$.   The density (\ref{neso}) satisfies the inhomogeneous Liouville
equation  
\[
\frac{\d \rho_{\epsilon} }{\d t}  \, + \, L \rho_{\epsilon} \, = \, 
             - \epsilon [ \,  \rho_{\epsilon} - \rhotilde  \,   ]     \, .   
\]
This equation shows that the nonequilibrium statistical operator $\rho_{\epsilon}$ becomes a
formal solution of the Liouville equation in the limit as $\epsilon \rightarrow 0+$.  
The usual procedure  to obtain 
a closure in terms of the mean resolved vector $a(t) $
is to impose the following two equations:
\begin{equation}   \label{zubarev}
\frac{d a }{ dt }  \, = \, \la \, LA \, | \, \rho_{\epsilon} \, \ra \, , \;\;\;\;\;  \mbox{ and } \;\;\; \;\;\;
a(t) = \la \, A \, | \, \rho_{\epsilon} (\cdot, t) \, \ra =    \la \, A \, | \, \rhotilde (\, \cdot \, ;\lambda(t)) \, \ra \, ,
\end{equation}  
for sufficiently small $\epsilon>0$ (or in an appropriate limit as $\epsilon \rightarrow 0+)$.    
The first equation, which is 
precisely the $A$-moment of the Liouville equation itself,  defines the
closed reduced equations for the macrostate $a$;  the second equation
is a consistency condition between $\rhotilde$ to $\rho_{\epsilon}$ that
determines $\lambda$ in terms of a common mean resolved vector $a$.    
Since $\rho_{\epsilon}$ relies on the memory of $A$ under the phase flow,
the resulting closure is governed by integro-differential equations in $a$, 
in which the propagator for the Liouville equation appears explicitly.       
 
The main justification for the closure theory based on nonequilibrium statistical operators
is that an irreversible thermodynamic formalism is achieved.     In this theory the irreversiblity
of the closed reduced dynamics for $a$ is produced by the slowly fading memory 
of the density $\rho_{\epsilon}$ used to form the instantaneous $A$-moment equations.    
That is, the use of $\rho_{\epsilon}$ rather than $\rhotilde$ breaks the time-reversal
symmetry of the adiabatic closure.    The reduction obtained in this way, however, is
nearly an identity, as opposed to a systematic method of approximating the influence 
of the unresolved variables on the resolved variables.    Implementation of the reduced
equations requires that ensembles of trajectories of the underlying Hamiltonian system
be computed over a time interval over which the memory fades.     In situations in which
there is a wide separation of time scales between the resolved and unresolved variables,
a further analysis is required to deduce the autonomous ordinary differential equations that
govern the macrostate.       

In these respects, the theory of nonequilibrium statistical operators shares the main features
of the well-known projection methods of nonequilibrium statistical mechanics
\cite{Balescu,Zwanzig,Robertson,Ottinger}.    In that
methodology, a stochastic integro-differential equation is produced by projecting
the Liouville propagators onto the resolved variables.   The result is an exact identity that
includes a time-convolution with a memory kernel, and random forcing terms that 
are related to the kernel via the fluctuation-dissipation theorem.   At least in the near-equilibrium
regime, the outcome of either theory is essentially the same, as are the challenges faced 
when implementing the reduced equations in a specific problem.   

For these reasons we pursue a fundamentally different approach in the present work.
In our approach the quasi-equilibrium densities, $\rhotilde(\lambda)$, are retained to
establish the thermodynamic structure of the reduced dynamics, but
moment closure is replaced by an optimization procedure over paths of trial densities,
$\rhotilde(\lambda(t))$.  The cost functional in this optimization is designed so
that the optimal path is that path which  
is most compatible with the underlying Hamiltonian phase flow in a
statistical, or information-theoretic,  sense.     
No other densities need be introduced to yield an irreversible closure,  nor is there a
need for a stochastic representation intermediate between the deterministic microscopic
dynamics and the macroscopic reduced equations.  

 The lack-of-fit cost function that is the basis of this procedure is constructed as follows.   
We introduce the  residual of  $\log \rhotilde$  with respect to the Liouville operator, namely,
\begin{equation}   \label{residual}
R(\, \cdot \,;\lambda,\lambdadot) \, = \, \left( \frac{\d}{\d t} + L \right) \log \rhotilde(\, \cdot \, ;\lambda(t)) \, .
\end{equation}    
[Here and throughout, $\lambdadot = d \lambda /dt$.]   
There are two related expressions for this statistic $R$, which we call the Liouville residual, 
that reveal its significance to the closure problem.    

First, for any $z \in \Gamma_n$
and any smooth parameter path $\lambda(t)$, the log-likelihood ratio 
between the density propagated for a short interval of time $\Delta t$ under
the Liouville equation  and the evolving trial density is 
\[
\log \frac{e^{-(\Delta t) L}  \rhotilde(z;\lambda(t)) }{ \rhotilde(z;\lambda(t+\Delta t))}
\; = \;  - (\Delta t) R(z;\lambda(t),\lambdadot(t)) \; + \; O ( \, (\Delta t )^2 \, )   
     \;\;\;\;\; \mbox{ as } \;\; \Delta t \rightarrow 0 \, .   
\]
This expansion shows that,  to leading order locally in time, $-R(z;\lambda(t),\lambdadot(t))$  represents the information in the sample point $z$ for discriminating the exact density 
against the trial density  \cite{CT,Kullback}.
By considering arbitrary smooth paths passing through a point $\lambda$ with 
a tangent vector $\lambdadot$,  we may consider $R(z;\lambda,\lambdadot)$ 
evaluated at $z$ as a function of $(\lambda,\lambdadot)$ in the tangent space
to the configuration space, and we may interpret it to be the local rate of information loss in
the sample point $z$ for the pair $(\lambda,\lambdadot)$.     
 
 Second, for any  dynamical
variable $F: \Gamma_n \rightarrow \R$, the $F$-moment of the Liouville
equation with respect to the trial densities along a path $\lambda(t)$ is
\[
\frac{d}{dt} \la F \, |\,  \rhotilde(\lambda(t)) \ra - \la LF \, | \,   \rhotilde(\lambda(t)) \ra
 \,=\,  \la FR \, |\,   \rhotilde(\lambda(t)) \ra \,  .
\]
Thus,  while an exact solution $\rho(t)$ of the Liouville equation satisfies
 (\ref{moment-eqn}) for all $F$, a trial solution $\rhotilde(\lambda(t))$ produces
a departure that coincides with  the covariance between
$F$ and $R$.   This representation  of $R$ furnishes a natural linear
structure for analyzing the fit of the trial densities to the Liouville equation.

In light of these two interpretations of the Liouville residual $R$,  we quantify the
dynamical lack-of-fit of the statistical model in terms it.    To do so, we consider the
components of $R$ in the resolved and unresolved subspaces.  At any configuration point
$\lambda \in \R^m$, let $P_{\lambda}$ denote the orthogonal projection of 
the Hilbert space $L^2(\Gamma_n, \rhotilde(\lambda))$  onto the span of the score functions
 $U_1(\lambda), \ldots , U_m(\lambda)$, 
and let $Q_{\lambda} = I - P_{\lambda}$ denote 
the complementary  projection; specifically,  for any $F \in L^2(\Gamma_n, \rhotilde(\lambda))$,  
\[
P_{\lambda} F = 
    \la F U(\lambda)^* \, | \,  \rhotilde(\lambda) \ra \,C(\lambda)^{-1}  U(\lambda)  \, . 
\]
where $C(\lambda)$ is given in (\ref{fisher}).  
We declare the lack-of-fit cost function to be 
\begin{equation}  \label{cost-fn}
\L (\lambda, \dot{\lambda})  \,=\, \half   \la  \,  (P_{\lambda} R)^2 \, | \, \rhotilde(\lambda) \, \ra
           + \half \la \, ( W_{\lambda} Q_{\lambda} R)^2 \, | \, \rhotilde(\lambda) \, \ra  \, .
\end{equation}
$\L$ is a weighted, mean-squared norm, with respect to a linear operator 
$W_{\lambda}$ on $L^2(\Gamma_n, \rhotilde) $ that assigns relative weights to the resolved and
unresolved components  of the Liouville residual.   
 $W_{\lambda}$ is assumed to be self-adjoint, and to satisfy
 $W_{\lambda}P_{\lambda} = P_{\lambda}W=P_{\lambda}$.   
It follows that the resolved and unresolved subspaces are invariant under  $W_{\lambda}$,
and that unit weights are assigned to the resolved subspace.   
The weights assigned by $W_{\lambda}$ to the unresolved subspace
constitute the adjustable parameters in the closure theory.    

A more explicit formula for  $\L$ is derived as follows.  The Liouville residual has
zero mean, $\la R | \rhotilde(\lambda) \ra=0$, and 
its orthogonal components are given by
\[
P_{\lambda} R = [\, \lambdadot - C(\lambda)^{-1} 
       \la \, L U(\lambda) \, | \, \rhotilde(\lambda) \, \ra \, ]^*U(\lambda) \, , 
\;\;\;\;\;\;\;\;   Q_{\lambda} R =  Q_{\lambda} L  \log \rhotilde(\lambda)  \, .
\]
The calculation of $P_{\lambda} R$ employs the  string of equations,
\begin{eqnarray*}
\la [L  \log \rhotilde(\lambda)] \, U(\lambda) \,|\, \rhotilde \ra &=& 
\int_{\Gamma_n}  \{ \rhotilde(\lambda) , H \} \, U(\lambda) \, dz   \\
   &=&   -    \int_{\Gamma_n}  \{ U(\lambda) , H \} \,  \rhotilde(\lambda) \, dz  \\ \nonumber
   &=&     - \la  L U(\lambda) \,| \, \rhotilde \ra  \, .    \nonumber 
\end{eqnarray*}
In light of these formulas the lack-of-fit cost function can be written in the form, 
\begin{equation}   \label{lagrangian}
\L (\lambda, \dot{\lambda})  \,=\, 
  \frac{1}{2} [\, \lambdadot - C(\lambda)^{-1} f(\lambda)  ]^*C(\lambda)
                       [\, \lambdadot - C(\lambda)^{-1}  f(\lambda) \, ]
     \;+\;  w(\lambda) \, , 
\end{equation}
where, compatibly with (\ref{f}), we employ the vector field
\begin{equation}   \label{f-general}
f(\lambda)  \, = \,    \la L U(\lambda) \, | \, \rhotilde(\lambda) \ra \, , 
\end{equation}  
and we define the non-negative function   
\begin{equation}  \label{w-general}
w(\lambda) = \half 
   \la [W_{\lambda} Q_{\lambda} L \log \rhotilde(\lambda)]^2 \, | \, \rhotilde(\lambda) \ra \, . 
\end{equation}   

We now formulate the stationary version of the optimization principle
that defines our statistical closure.    For an initial time, $t_0$, we consider
the dynamical minimization problem
\begin{equation}    \label{value-fn}
v(\lambda_0) \, = \,   \min_{\lambda(t_0)=\lambda_0} \int_{t_0}^{+ \infty}  
        \L(\lambda, \lambdadot)  \, dt \,  , \;\;\;\;\;\;\;\;\;\;
  \mbox{ for }  \;\;  \lambda_0 \in \R^m \, ,  
\end{equation}    
in which the lack-of-fit cost function (\ref{lagrangian}) defines the cost functional
over admissible paths $\lambda(t), \; t_0 \le t < + \infty \, $, in the configuration space of the
statistical model, with the constraint that each path starts at $\lambda_0$ at time
$t_0$.    In conformity with optimization and control theory, we refer to $v(\lambda_0)$
as the value function for the minimization problem (\ref{value-fn}) \cite{BH,FR,GF}.    Since 
the cost function $\L(\lambda, \lambdadot)$ is independent of $t$, and the optimization
extends to infinity in time, $v(\lambda_0)$ is independent of $t_0$;  indeed, the
time variable in (\ref{value-fn}) can be shifted so that $t_0$ is replaced by $0$.    
It is in this sense that we refer to (\ref{value-fn}) as the stationary formulation
of the best-fit principle.

By analogy to analytical mechanics, one may regard (\ref{value-fn}) as a principle of
least action for the ``Lagrangian" $\L(\lambda, \lambdadot)$ and interpret the
first member  in (\ref{lagrangian}) as 
its ``kinetic" term and the second member as its ``potential" term \cite{Arnold,Lanczos}.
The kinetic term is a quadratic form in the generalized velocities $\lambdadot$ with positive-definite
Fisher information matrix $C(\lambda)$, and it is entirely determined by the expressions
that also arise in the adiabatic closure presented in the previous section.   
The potential term $w(\lambda)$ embodies the influence of the unresolved
variables on the resolved variables, and it depends on the weight operator $W_{\lambda}$
that quantifies that influence.     Accordingly, we call $w(\lambda)$ the closure potential.    
 Of course, these mechanical analogies are not literal: 
 $\L$  has the units of a rate of entropy production, not of an energy, and  
it is a sum of its ``kinetic" and ``potential" terms, not a difference.         

An extremal path $\lambdahat(t)$, with $\lambdahat(t_0) = \lambda_0$, for (\ref{value-fn}) determines
the best-fit evolution of the statistical macrostate $\rhotilde(\, \cdot \, ; \lambdahat(t))$,  
in the sense that the time-integral of the lack-of-fit cost function $\L$ is minimized 
along the path $\lambdahat(t)$.    Thus, the extremal paths for  (\ref{value-fn}) define the
estimated, or predicted, evolution in our reduced model, and the closed equations governing
this evolution are derived from the optimization principle  (\ref{value-fn}).    
 
The derivation of the closed reduced equations that follow from the optimality conditions
for (\ref{value-fn}) makes use of Hamilton-Jacobi theory \cite{Arnold,Lanczos,GF,Evans}.     
The value function, $v(\lambda)$, on $\lambda \in \R^m$, defined by (\ref{value-fn}), 
is analogous to an action integral, or Hamilton principal function, and it therefore satisfies the
associated time-independent Hamilton-Jacobi equation,
\begin{equation}  \label{HJ-stat}
\H \left( \lambda, \, - \frac{\d v}{\d \lambda} \right) \, = \, 0 \, ,
\end{equation}
where $\H(\lambda, \mu)$ is the Legendre transform of $\L(\lambda, \lambdadot)$.  
That is,  $\L$ and $\H$ are convex conjugate functions with respect to their second
arguments, and 
\begin{eqnarray}  \label{legendre}
\mu &= &   \frac{ \d \L }{\d \lambdadot} \, = \, C(\lambda)  \lambdadot - f(\lambda)   \\  \nonumber
\H(\lambda, \mu) &=&  \lambdadot^*\mu - \L(\lambda,\lambdadot) \, = \, 
     \half \mu^* C(\lambda)^{-1} \mu +  f(\lambda)^*C(\lambda)^{-1} \mu  - w(\lambda) \, .   
\end{eqnarray}
These calculations are straightforward and explicit because $\L(\lambda,\lambdadot)$
is a quadratic function of $\lambdadot$.     According to Hamilton-Jacobi theory, 
the conjugate variable $\mu = \hat{\mu}(t)$ along an extremal
path $\lambda = \lambdahat(t)$ is given by the relation 
\begin{equation}   \label{conjugate-rel}
\hat{\mu} \, = \,   - \frac{\d v}{\d \lambda}(\lambdahat)      \, . 
\end{equation}
This basic relation
together with (\ref{legendre}) closes the reduced dynamics along the extremal path,  in
that it yields the (vector) differential equation 
\begin{equation}  \label{closed-reduced}  
C(\lambdahat) \, \frac{d \lambdahat}{dt}  \, = \, f(\lambdahat)  \, - \, \frac{\d v}{\d \lambda}(\lambdahat)  \, . 
\end{equation}   

In summary, the choice of a statistical model of trial probability densities $\rhotilde(\lambda)$ and
the specification of a weight operator $W_{\lambda}$ determine the lack-of-fit Lagrangian
$\L$ and its associated Hamiltonian $\H$, and thereby the value function $v(\lambda)$ 
for the best-fit optimization principle.    All the terms in (\ref{closed-reduced}) are then uniquely
defined, and this equation governs the evolution of best-fit statistical states in the reduced model.   
The modeled relaxation is described by an extremal path $\lambdahat(t)$ 
satisfying  $\lim_{t \rightarrow +\infty} \lambdahat(t) \, = 0 \, $, since 
the trial densities $\rhotilde(\lambda)$ are parameterized so that 
$\rhotilde(0)$ is an equilibrium density.   

The value function satisfies the equilibrium conditions 
\begin{equation}  \label{equil-cond}
v(0) =0 \, , \;\;\;\;\;\;\;\;  \frac{\d v}{\d \lambda}(0) \, = \, 0 \, ; 
\end{equation} 
the first is immediate  from (\ref{value-fn}), and the second is a direct consequence
of   (\ref{HJ-stat}).     Moreover, the Hessian 
matrix of second partial derivatives of the value function is non-negative, semi-definite at equilibrium,
\[
\xi^* \frac{\d^2 v}{\d \lambda \d \lambda^*}(0) \xi \, \ge 0 \, ,   \;\;\;\;\;\;\;
  \mbox{ for all } \; \xi \in R^m \, .    
\]
In the non-degenerate case when the resolved variables do not contain any conserved
quantities, this matrix is actually positive-definite.   These properties reflect that fact that 
the equilibrium state $\lambda=0$ is a local minimum of $v(\lambda)$, and in the non-degenerate case 
a strict local minimum.   
The value function defined by (\ref{value-fn}) 
is the so-called viscosity solution of (\ref{HJ-stat}) validating these equilibrium conditions.
The reader is referred to \cite{Evans} for the modern theory of existence and uniqueness
of solutions to Hamilton-Jacobi equations, including the time-dependent equations  
encountered in Section 5.

The foregoing discussion applies to the most general form of the reduction using
any regular family of densities $\rhotilde(\lambda)$ as the statistical model.    We now
 specialize the results contained in (\ref{closed-reduced}) 
 to the models that use quasi-equilibrium densities (\ref{canonical-density}), for
 either  fixed inverse temperature $\beta$ or  fixed mean energy $E$.      

For the family (\ref{qe-density}) with fixed $\beta >0$, the lack-of-fit Lagrangian takes the form
(\ref{lagrangian}) with  $C(\lambda) = \la (A-a) (A-a)^* \, | \, \rhotilde(\lambda) \ra$ and 
$f(\lambda) = \la LA \, | \, \rhotilde(\lambda) \ra$.   The closure potential can be expressed
in the explicit form  
\begin{equation}   \label{w-beta}
  w(\lambda) \, = \,   \half 
     \lambda^* \la \, [ W_{\lambda} Q_{\lambda} LA ] \, [W_{\lambda} Q_{\lambda} LA^* ] \, | 
                                  \, \rhotilde(\lambda)  \, \ra  \, \lambda \, 
                        = \, \half   \lambda^* D(\lambda)  \lambda \, , 
\end{equation}  
where the second equality defines the $m \times m$ Gram matrix $D(\lambda)$.     
Thus, $w(\lambda)$ is expressed as a non-negative quadratic form in $\lambda$, with
a symmetric, semi-definite matrix $D(\lambda)$ which, in general, is $\lambda$-dependent.    
It is important to note that the weight operator $W_{\lambda}$ enters into the best-fit
closure scheme only through the matrix $D(\lambda)$, and so all the adjustable parameters
introduced into the scheme as weight factors in the lack-of-fit cost function are assembled
in  $D(\lambda)$ alone.    The practical utility of this approach to closure rests
on the requirement that in problems of interest there be a natural choice of
weights via $W_{\lambda}$ that leads to a  tractable number 
of adjustable parameters in $D(\lambda)$ and hence in $w(\lambda)$.    

The closed reduced equations governing the evolution of the best-fit macrostate
are 
\begin{equation}   \label{closed-reduced-beta}
\frac{d \ahat}{dt} \, = \, f(\lambdahat)  \, - \,  
                             \frac{ \d v}{\d \lambda}(\lambdahat)    \, ,  \;\;\;\;\;\; \;\;
         \mbox{ with } \;\;\;\;\;   \ahat(t) \, = \, \la A \, | \, \rhotilde(\lambdahat) \ra  \, . 
\end{equation}
Referring to (\ref{adiabatic-eqn}) we see that (\ref{closed-reduced-beta}) differs
from the adiabatic reduced equation derived in the previous section by the presence
of the gradient of $v(\lambda)$.    In the subsequent section, this gradient term is shown to 
be the source of entropy production in the closed reduced dynamics, and consequently
the best-fit evolution governed by (\ref{closed-reduced-beta}) is irreversible.    
It is illuminating to display the Hamilton-Jacobi equation satisfied by $v(\lambda)$,
namely,
\begin{equation}    \label{HJ-beta}
\half \left( \frac{\d v}{\d \lambda}  \right)^* C(\lambda)^{-1} \left( \frac{\d v}{\d \lambda}  \right)
\, - \, f(\lambda)^* C(\lambda)^{-1}\left( \frac{\d v}{\d \lambda}  \right) \, = \, 
    \half \lambda^* D(\lambda) \lambda \, .  
\end{equation}
The source term in this partial differential equation for $v(\lambda)$ is the closure potential
with matrix $D(\lambda)$.  This matrix quantifies the magnitude of the unresolved Liouville residual
and, in turn, generates the irreversible part of the closed reduced equation for $\ahat(t)$  
via (\ref{HJ-beta}).   In this way  the irreversible reduced macrodynamics
ensues from the reversible, conservative microdynamics.      

The same governing equations hold for the version of the reduction in which the
mean energy $\la H \, | \, \rhotilde \ra = E$ is fixed.   Specifically, the closed
reduced equation governing this version is identical to (\ref{closed-reduced-beta}),
the only change being that $C(\lambda)$ is replaced by the appropriate matrix
(\ref{fisher-energy});  the terms $f(\lambda)$ and $w(\lambda)$ take the same forms 
as in the version with fixed $\beta$.  

 The reversible term in either of these reduced equations  is expressible as a Hamiltonian
vector field in the same way as in Section 2.  That is, the
effective Hamiltonian for the version with fixed $\beta$ is 
the free-energy function, $\tilde{h}$, while for fixed $E$ it is the mean energy function, $h$.

\section{Properties of the reduced model  }

The best-fit statistical closures developed in the preceding section result in 
reduced equations for the macrostate $a$ which possess 
the so-called GENERIC (General Equations of NonEquilibrium Reversible-Irreversible
Coupling) format, a general framework for the equations of thermodynamics
and hydrodynamics \cite{GO,OG,Ottinger-book}.
Let us consider the formulation with a fixed 
mean energy $ \la H \, | \, \rhotilde(\lambda) \ra = E $, for which the reversible dynamics
is given by (\ref{adiabatic-ham-E}).    The governing equation (\ref{closed-reduced}) 
for the best-fit closure has both reversible and irreversible terms, which are expressible
in the form   
\begin{equation}    \label{generic-E}  
\frac{d \ahat}{dt} \, = \,  \Omega(\ahat)  \, \frac{ \d h}{\d a}(\ahat)  \, - \, 
     \frac{ \d v}{\d \lambda}(\lambdahat)    \, , \hspace{2.5cm}
     \lambdahat = - \frac{\d s}{\d a} (\ahat)   \, .   
\end{equation}  
As in Section 2, $s=s(E,a)$ denotes the entropy relation $s=s(u,a)$ in (\ref{max-ent}) 
evaluated at fixed energy, $E$, 
and the mean energy function, $u=h(s,a)$, is defined by inverting this relation.
 The reversible term in (\ref{generic-E})
is a Hamiltonian vector field with Hamiltonian $h$ and cosympletic matrix, $\Omega$;
the gradient $\d h / \d a$ is at fixed entropy.    The irreversible
term  in (\ref{generic-E}) is a gradient vector field of the value function $v$ with respect to 
$\lambda= -\d s / \d a$.    This form of the irreversible term conforms precisely
with the general, nonlinear GENERIC form proposed by Grmela and {\"O}ttinger \cite{GO,OG}.  
Other closely related formats have also been developed \cite{Morrison,BE,Berdichevsky}.     

By virtue of  (\ref{conjugate-rel}), the irreversible term in (\ref{generic-E}) is exactly 
the conjugate vector $\hat{\mu}$.  
 This correspondence furnishes the  thermodynamic
interpretation of the vectors $\lambda$ and $\mu$ in our optimization-based theory.
Namely,
the variables $ \lambda^1, \ldots,  \lambda^m$, which are the ``coordinates" of the
Hamilton-Jacobi theory, may be interpretated as (minus) thermodynamic forces,
or affinities;  and,  the variables $\mu_1, \ldots , \mu_m$, which are the conjugate 
``momenta", may be interpreted as  corresponding thermodynamic fluxes.     
[Our notation introduces a minus sign in $\lambda$ compared to much of
the physical literature.]    Thus, the duality  naturally attached to our variational 
principle (\ref{value-fn}) coincides with the duality between ``forces" and ``fluxes"  
in nonequilibrium thermodynamics \cite{dGM,Keizer}.   

The value function, $v(\lambda)$, which mediates this duality, 
is closely related to the entropy production.  Namely, an immediate
implication of  (\ref{generic-E}) is the fundamental identity, 
\begin{equation}  \label{entropy-prod}
\frac{d \hat{s}}{dt} \, = \, - \lambdahat^*\hat{\mu} \, = \, 
             \lambdahat^*\frac{\d v}{\d \lambda} (\lambdahat) \, . 
\end{equation}
An entropy production inequality follows from this identity, provided
that $v(\lambda)$ is convex.    This convexity is guaranteed whenever the cost
function $\L(\lambda,\lambdadot)$ is jointly convex in the pair $(\lambda,\lambdadot)$;  but,
for large $|\lambda|$ the dependence of the coefficient functions $C(\lambda)$ and $D(\lambda)$,
as well as the nonlinearity of $f(\lambda)$, may result in a loss of convexity.   
Nonetheless, the convexity of $v(\lambda)$ always holds in some
 neighborhood of equilibrium, $\lambda=0$.   
Thus, apart from a far-from-equilibrium regime in which convexity may be lost,
we are assured that entropy increases at a rate bounded below by $v$:   
 \begin{equation}  \label{second-law}
\frac{d \hat{s}}{dt} \, \ge \,  v( \lambdahat) \,  \ge 0 \, .  
\end{equation} 
This inequality follows from the fact that $0 = v(0) \ge v(\lambda) - \lambda^*\d v / \d \lambda (\lambda)$
for $\lambda$ in the domain of convexity of $v$.       
The value function is thus found to be the key thermodynamic potential in the nonequilibrium
reduced model, and its close relation to entropy production further reinforces the 
information-theoretic basis of the optimization principle (\ref{value-fn}).   

Since nonequilibrium behavior is most readily understood in the near-equilibrium regime,
we now present the best-fit closure theory in its approximate form for small $|\lambda|$; that is,
for paths $\lambda(t)$ in a neighborhood of  $\lambda=0$ having trial densities
$\rhotilde(\lambda(t))$ close to $\rho_{eq}$.        
We assume that the resolved vector $A$ is normalized so that
$\la A \ra_{eq} = 0$.     We also assume that $\la A H \ra_{eq} =0$;
this is easily obtained by replacing each $A_i$ by $A_i - \alpha_i (H-E)$ for appropriate $\alpha_i$.   
Under these normalizations the linearization of the closure is identical for the
versions  that fix either $\beta$ or $E$ along paths.   

The near-equilibrium form of the cost function (\ref{lagrangian}) is simply
\begin{equation}  \label{lagrangian-ne}  
\L(\lambda, \lambdadot) \, = \, 
  \half [ \, \lambdadot - C^{-1} J \lambda \, ]^* C  [ \, \lambdadot - C^{-1} J \lambda \, ]
   + \half \lambda^*D\lambda \, , 
\end{equation}   
in which the coefficient matrices are constant, being evaluated at equilibrium; specifically,
\begin{eqnarray*}
C =C(0) = \la A A^* \ra_{eq} \, , \;\;   D =D(0) = \la (WQLA)(WQL A^*) \ra_{eq} \, , \;\; 
J = \frac{\d f}{\d \lambda} (0) = \la (LA) A^* \ra_{eq} \, .   
\end{eqnarray*}
$C$ is a positive-definite symmetric matrix, $D$ is a semi-definite symmetric matrix,
and $J$ is an anti-symmetric matrix.     
The cost function (\ref{lagrangian-ne}), which is a quadratic form in $(\lambda,\lambdadot)$, 
governs the linear relaxation from near-equilibrium initial states $\lambda_0$,
a phenomenon treated by familiar linear-response theory \cite{Chandler,Zwanzig}.       
The value function associated with $\L$ is therefore also a quadratic form, 
and in light of (\ref{equil-cond}) it is simply
\begin{equation}    \label{v-ne}
v(\lambda) = \half \lambda^* M \lambda \, , 
\end{equation}   
for some constant, symmetric $m \times m$ matrix $M$.   
The Legendre transform yields the following  expressions
 for the conjugate vector of fluxes and  the Hamiltonian:
\[
\mu = C\lambdadot - J \lambda \, , \;\;\;\;\;\;\;
\H(\lambda,\mu)= \half \mu^*C^{-1} \mu  - \lambda^*J C^{-1} \mu - \half \lambda^* D \lambda \, . 
\]
Along an extremal path, $\lambdahat(t)$, the relation between flux and force is linear, 
$\hat{\mu} = - M \lambdahat$.   Consequently, the closed reduced equation for
near-equilibrium relaxation is the constant-coefficient linear system of ODEs, 
\begin{equation}   \label{closed-reduced-ne}
\frac{d \ahat} {dt} \, = \, (  J - M  )  \, \lambdahat \, ,  \;\;\;\;\;\;\;\;\;\;\;\;\;\;
                       \ahat = C \lambdahat     \, .   
\end{equation}
The matrix, $M$, is determined by an algebraic Riccati equation,
to which the stationary Hamilton-Jacobi equation (\ref{HJ-stat}) reduces \cite{Kucera}; 
namely,  
\begin{equation}   \label{riccati-stat}
M C^{-1} M + J C^{-1} M - M C^{-1} J \, = \, D \, .    
\end{equation}
The coefficient matrices, $C$ and $J$, are determined entirely by the parametric statistical
model, whereas the source matrix, $D$,  depends on the choice of the adjustable 
constants in the closure.  $M$ is the unique, semi-definite matrix solution of (\ref{riccati-stat}).
The matrix of transport coefficients in (\ref{closed-reduced-ne}) thus includes an
antisymmetric (reversible) part and a symmetric (irreversible) part, and while
the irreversible part, $-M$  responds directly to the choice of closure matrix $D$,
it is also modified by the reversible part, $J$, via the coupling terms in the Riccati equation.  

Under the near-equilibrium approximation the entropy production identity (\ref{entropy-prod})
becomes 
\[
\frac{d \hat{s}}{dt} \, = \,  2v( \lambdahat) \, = \,  
    \int_t^{+\infty} 2 \L(\lambdahat(t'), d\lambdahat/dt') \, dt'  \;    \ge 0 \, ,  
\]
since $v$ is homogeneous of degree 2, and hence $\lambda^*\d v / \d \lambda = 2 v$.   
Thus the time-integrated, squared-norm of the lack-of-fit along the extremal $\lambdahat$
equals the entropy production of the predicted evolution.   In this light it is evident
that the dynamical optimization principle defining our best-fit closure theory establishes   
the tight relationship between the rate of information loss in the reduced
dynamics and the quantified lack-of-fit of the trial densities to the Liouville equation.        
The one-to-one correspondence between $M$ and $D$ 
in (\ref{riccati-stat}) expresses this relationship for relaxation in a neighborhood of equilibrium.

\section{Nonstationary formulation  }

The stationary formulation of the best-fit closure developed in the preceding sections
has an irreversible term that is time-independent, being the minus gradient of the
solution $v(\lambda)$ of the time-independent Hamilton-Jacobi equation (\ref{HJ-stat}).   
But when the modeled evolution is initiated by a quasi-equilibrium density 
$\rhotilde(\lambda_0)$ for a given initial state $\lambda_0 \in \R^m$,
there is an early phase of the relaxation in which the entropy production increases
from zero.   After a sufficiently long time, the entropy production and the irreversible
flux approach their stationary values.  This ``plateau" effect can be demonstrated by 
expanding an exact solution of the Liouville equation in a power series in time about
$t= 0$.    For small $t$, the mean resolved vector corresponding to the exact solution 
is given by 
\[
a(t) \, = \, \la \, e^{t L} A \, | \, \rhotilde(\lambda_0)  \ra 
    \, = \, \la A \, | \,   \rhotilde(\lambda_0) \ra \, + \, t  f(\lambda_0) 
        \, + \, O(t^2) \, , 
\] 
recalling the definition of $f(\lambda)$ in (\ref{f}).      
If we attach to $a(t)$ a path of trial densities, $ \rhotilde(\lambda(t))$, with $\lambda(t)$ 
determined so that  $a(t) = \la A \, | \,  \rhotilde(\lambda(t)) \, \ra$, then 
\[
\lambda(t) \, = \, \lambda_0 \, + \,  t C(\lambda_0)^{-1} f(\lambda_0) \, + O(t^2) \, . 
\]
The entropy production along this path is then
\[
\frac{d s}{dt}(a(t)) \, = \,  -\lambda(t)^*\frac{da}{dt}(t) \, = \, 
      \lambda_0^*f(\lambda_0) \, + \, O(t)  \, ;    
\]
and the constant term in this equation vanishes,
\[
  \lambda_0^*f(\lambda_0) \, = \, \la L(\lambda_0^*A) \, | \, \rhotilde(\lambda_0) \ra \, = \, 
     \int_{\Gamma_n} L  \rhotilde(\lambda_0) \, dz  \, = \, 0  \, .   
\]
Thus, we see that the entropy production of a perfectly fitted path of trial densities grows from
zero linearly in time, given an initial density that is itself a trial density.    

Physically, such an initial condition is naturally created by applying constant 
external perturbations with strengths proportional to  $\lambda_0^1, \ldots , \lambda_0^m$  
via the resolved variables and allowing the system to equilibriate 
with respect to the perturbed Hamiltonian $H - \beta^{-1} \lambda_0^*A $.     
When  these external perturbations are switched off at time $t=0$, a relaxation occurs
that is the object of the modeling exercise.    Moreover, in numerical experiments 
that test a closure theory it is often convenient to use initial ensembles
defined by quasi-equilibrium trial densities.     
For these reasons,  we now  modify
the defining optimization principle (\ref{value-fn}) to include such a plateau effect.    
We refer to this 
as the nonstationary formulation, in that the value function $v =v(\lambda,t)$ becomes
time dependent.   

The nonstationary optimization principle has the value function   
\begin{equation}    \label{value-fn-non}
v(\lambda_1,t_1) \, = \,   \min_{\lambda(t_1)=\lambda_1} \; \int_{0}^{t_1}  
        \L(\lambda, - \lambdadot)  \, dt \,  , 
 \end{equation}   
in which the admissible paths 
$\lambda(t), \; 0 \le t \le t_1 \, $, in the configuration space of the
statistical model are constrained to terminate  at $\lambda_1 \in \C^n$ at time $t_1 \ge 0$,
while $\lambda(0)$ is unconstrained.   The integrand in (\ref{value-fn-non}) is modified 
to account for time-reversal, and the admissible paths may be viewed as
evolving in reversed time,
$\tau = t_1 -t$, starting from $\lambda_1$ at $\tau=0$.   The value function $v(\lambda_1, t_1)$
quantifies the optimal lack-of-fit of a time-reversed path connecting 
 the current state $\lambda_1$ to an unspecified initial state $\lambda(0)$.   
By contrast, the stationary formulation optimizes over admissible paths in forward time 
that join the current state $\lambda_0$ at time $t_0$ to unspecified limit states 
as $t \rightarrow +\infty$; convergence of the time integral then requires that the extremal
tends to equilibrium.        
In essence, the nonstationary formulation ties the value function to the time at which
 the initial condition is specified, while the stationary formulation ties the value function
to equilibration at infinite time.        

 The value function (\ref{value-fn-non}) is the unique solution of the initial value
problem (with $(\lambda,t)$ replacing $(\lambda_1,t_1$) )
\begin{equation}  \label{HJ-non}
 \frac{\d v}{\d t}  \, + \,  \H \left( \lambda, \, - \frac{\d v}{\d \lambda} \right) \, = \, 0 \, , \;\;
    \mbox{ for } \; t>0,  
    \;\;\;\;\; \mbox{ with } \;   v(\lambda,0)=0 \, ,  
\end{equation}
which is a time-reversed Hamilton-Jacobi equation.  That is, 
keeping $\H(\lambda, \mu)$ the Legendre transform of $\L(\lambda, \lambdadot)$ 
as in (\ref{legendre}), the Hamiltonian corresponding to $\L(\lambda, -\lambdadot)$     
is    $\H(\lambda, - \mu)$; thus (\ref{HJ-non}) follows from (\ref{value-fn-non}).      
 
In terms of the time-dependent value function, $v(\lambda,t)$, closure is achieved 
by setting 
\begin{equation}  
\hat{\mu} (t) \, = \, - \frac{ \d v}{ \d \lambda } (\lambdahat, t)   \, .
\end{equation}   
This relation is the nonstationary analogue to (\ref{conjugate-rel}).  
The nonstationary reduced equations 
are determined by this relation in the same way that
(\ref{closed-reduced}) follows from (\ref{conjugate-rel}); namely,
\begin{equation}   \label{closed-reduced-non}  
C(\lambdahat) \, \frac{d \lambdahat}{dt}  \, = \, f(\lambdahat)  \, - \, \frac{\d v}{\d \lambda}(\lambdahat,t)  \, . 
\end{equation}    
But, in contrast to the stationary case, the solution $\lambdahat(t)$ of (\ref{closed-reduced-non}) 
is not itself an extremal of the defining optimization principle; the extremal paths that
define $v(\lambda,t)$  for each $(\lambda,t)$ vary with $t$ as well as with $\lambda$,
and they instantaneously determine the flux of the predicted evolution $\lambdahat(t)$.     

Since $\H(\lambda,\mu)$ is positive-definite in $\mu$, the viscosity  
solution $v(\lambda,t)$
of (\ref{HJ-non}) exists for all time $t>0$,  is unique,  and  coincides with
the value function defined by (\ref{value-fn-non}) \cite{Evans}.       
Moreover,   $v(\lambda,t) \rightarrow v(\lambda)$, the stationary value function, as $t \rightarrow +\infty$. 
Thus, the nonstationary best-fit closure is a natural generalization 
of the stationary closure that straightforwardly includes an intrinsic plateau effect.   

We remark that it is possible to consider solutions of   (\ref{HJ-non}) having
nonzero initial conditions $v(\lambda,0)$.   One natural alternative is to optimize over
 time-reversed admissible paths $\lambda(t)$ that attach to
the given initial state, so that $\lambda(0)=\lambda_0$.    
This requirement produces a value function, $V(\lambda,t;\lambda_0)$, that is singular
at $t=0$ in the sense that  
$V(\lambda,t;\lambda_0) \sim (\lambda - \lambda_0)^*C(\lambda_0) (\lambda - \lambda_0)/ 2t $
as $t \rightarrow 0+$.    A theory based on this value function, and using (\ref{closed-reduced-non})
with $V$ replacing $v$, constitutes another natural formulation of the plateau effect.   
The resulting reduced equations, however, are less attractive as a tractable closure because
they are inhomogeneous in the mean resolved vector $\ahat$,  they require a more
intricate solution of (\ref{HJ-non}), and their properties are more difficult to deduce.    Consequently,
even though this approach incorporates the plateau effect in a strong form, we prefer
the  nonstationary theory that follows from the homogeneous initial condition $v(\lambda,0)=0$
in (\ref{HJ-non}), and that shares the general properties the stationary theory.

In particular, the entropy production identity (\ref{entropy-prod}) and the inequality (\ref{second-law})
remain valid in the homogeneous, nonstationary closure theory, since the time-dependent value function
$v(\lambda,t)$ continues to satisfy the equilibrium conditions (\ref{equil-cond}).   Moreover, the Hessian
matrix of second partial derivatives of the value function evaluated at equilibrium,
\[
M(t) \, = \, \frac{\d^2 v}{\d \lambda \d \lambda^*}(0, t) \,  ,
\]  
satisfies the  Riccati differential equation
\begin{equation}   \label{riccati-non}
\frac{dM}{dt} + M C^{-1} M + J C^{-1} M - M C^{-1} J \, = \, D \, ,   \;\;\;\;\;\;\;\;\;\;  M(0)=0 \, .    
\end{equation}
The derivation of (\ref{riccati-non}) merely requires taking the matrix of second partial
derivatives of (\ref{HJ-non}) and evaluating the result at $\lambda=0$.     
This equation implies that $M(t)$ is semi-definite, because
the source term $D$ is semi-definite; in the non-degenerate case when $D$ 
is positive-definite, which is the  case whenever none of the resolved variables
is a dynamical invariant, $M(t)$ is positive-definite for $t>0$ \cite{Kucera}.  
The  equilibrium point $\lambda=0$ is then a strict local minimum for $v(\lambda,t)$
for all $t>0$.

The near-equilibrium linearization of the nonstationary best-fit closure theory
has a time-varying linear relation between flux and force along an extremal path, 
 $\hat{\mu} = - M(t) \lambdahat$.   
Consequently, the nonstationary version of the closed reduced equation
 is a nonautonomous linear system of ODEs
entirely analogous to (\ref{closed-reduced-ne});  namely,
\begin{equation}   \label{closed-reduced-ne-non}
\frac{d \ahat} {dt} \, = \, [  J - M(t)  ]  \, \lambdahat \, ,  \;\;\;\;\;\;\;\;\;\;\;\;\;\;
                       \ahat = C \lambdahat     \, .   
\end{equation}
The value function $v(\lambda,t)$ is the time-dependent quadratic form (\ref{v-ne}).  

As a simple illustration let us consider the near-equilibrium, nonstationary closure for a
single resolved variable $A \in \R^1$, that is, for $m=1$.    Then $C>0$, $J=0$, and there is a single
closure parameter $D\ge0$.    The solution to the scalar Riccati differential equation (\ref{riccati-non})
and  the closed reduced equation (\ref{closed-reduced-ne-non}) are, respectively, 
\[
M(t) = \sqrt{CD} \tanh \sqrt{\frac{D}{C}}\,  t  \; ,
\;\;\;\;\;\;\;\;\;\;
a(t) \, = \, a_0 \, \mbox{sech}   \sqrt{\frac{D}{C}} \, t  \; . 
\]
This solution clearly shows the plateau effect for small $t$ and the exponential relaxation for
large $t$.      

The analogous multivariate form of this result holds for $m>1$ provided
that all the resolved variables $A_1, \ldots , A_m$, as well as $H$,  are even under time-reversal;
that is, $A_k(q,-p) = A_k(q,p)$, for $k=1, \ldots , m$.  Then, each $LA_k$ is odd, 
and consequently  $J = \la (LA) A^* \ra_{eq}=0$.    And the same holds if all the
$A_k$ are odd.   In these situations the closed reduced dynamics
governed by   (\ref{closed-reduced-ne-non}) and (\ref{riccati-non}) is diagonalizable.   
Namely, let $V$ be the $m \times m$ matrix of eigenvectors for $D$ relative to $C$; that is,
$V$ is the matrix for which $V^*CV=I$ and $V^*DV=\Delta$, where $\Delta$ is the diagonal
matrix whose diagonal consists of the associated eigenvalues 
$\gamma_1, \ldots , \gamma_m \ge 0$.      In the transformed variables, $b(t)=V^*a(t)$ and 
$N(t) =V^*M(t)V$, the closed reduced dynamics    (\ref{closed-reduced-ne-non}) and 
(\ref{riccati-non}) become 
\[
\frac{d \hat{b} }{dt} \, = \, - N(t) \,  \hat{b} \, , \hspace{1cm} \frac{dN}{dt} + N^2 \,= \, \Delta \, .   
\]   
The solution $N(t)$ of this transformed Riccati equation is the diagonal matrix
having diagonal elements,  $ \, \sqrt{\gamma_1} \tanh   \sqrt{\gamma_1} t , \ldots , 
\sqrt{\gamma_m} \tanh  \sqrt{\gamma_m} t  \,$.   In turn, the solution of the
relaxation equation is $\hat{b}(t) = ( \, b_1(0) \,  \mbox{sech} \sqrt{\gamma_1} t , \ldots, 
b_m(0) \, \mbox{sech} \sqrt{\gamma_m} t \, ) $.   In the original variables the relaxation
is given by $\ahat(t) = (V^*)^{-1}  \hat{b}(t)$ and $\lambdahat(t)= V^{-1}  \hat{b}(t)$.  
Thus, the best-fit closure theory for these resolved variables predicts a linear relaxation
that combines a plateau effect with an exponential decay for $m$ normal
modes with $m$ characteristic time scales $1/\sqrt{\gamma_k}$.      

In the general near-equilibrium relaxation for which special symmetries are not present
in the selected resolved variables, the antisymmetric matrix $J$ is nonzero, and it
participates in  the Riccati matrix equation (\ref{riccati-non}) that determines $M(t)$.  
Consequently,  $J$, which defines the reversible part of closed reduced dynamics,
also influences the irreversible part controlled by $M(t)$.   A central prediction 
of the best-fit closure theory, therefore, is the coupling between the reversible
and irreversible parts of the closed reduced equations, which implies a quantitative
coupling between the time scales of reversible oscillations and the time 
scales of irreversible relaxation.        

\newpage
\section{Discussion  }

We have developed a methodology  for constructing 
nonequilibrium statistical models of Hamiltonian dynamical systems 
based on the recognition that such models are essentially estimation strategies.  
When seeking closure in terms of a reduced description,
it is desirable to have the flexibility 
to expand or contract the set of resolved, or macroscopic, variables, while maintaining
a consistent approximation strategy.    Expansion increases accuracy at the
expense of model complexity, while contraction improves tractability at the
expense of model fidelity.   Whatever the compromise between accuracy and tractability,
the goal of the modeling exercise is to obtain a dynamical closure 
that is as effective as possible for the selected macrovariables.          
Motivated by these considerations,  
we have developed a best-fit criterion for canonical statistical models associated 
with any independent set of resolved variables.     

Our optimization principle produces an irreversible, dissipative macrodynamics
that minimizes a certain lack-of-fit of the selected statistical model to the reversible, 
conservative microdynamics.       The structure of the cost function in this principle
conforms to the separation of the reduced dynamics into reversible and irreversible
parts, in that it assigns relative weights to the resolved and unresolved components
of the Liouville residual.   In particular,
the contribution of the unresolved component to the cost function is 
the source of irreversibility in the best-fit closure, and the rates of relaxation of the 
macrovariables scale with the weights assigned to the unresolved component.      

The closed reduced equations that follow from the optimality conditions have
the form of governing equations for nonequilibrium thermodynamics.    The
value function, which is the central quantity in the optimization theory, becomes
the pivotal quantity in the thermodynamic formalism.    Specifically,  
the gradient of the value function with respect to the
vector of affinities is the vector of corresponding fluxes.   The
value function is therefore the natural extension of the Rayleigh-Onsager dissipation function
beyond the range of linear irreversible thermodynamics.    The main novelty of
our variational formulation is that this key thermodynamic potential is realized as 
the minimal lack-of-fit of macroscopic paths to the underlying microscopic dynamics.    
 
The utility of reduced models and statistical closures for any particular system
depends upon two factors: (i) a convenient set of the resolved variables 
with some separation of time scales between the the resolved (slow) and unresolved (fast)
variables; (ii) an expression for, or approximation of, the dissipation function
in terms of  a manageable number of adjustable parameters.   Traditionally, thermodynamic modeling
has tended to concentrate on special physical situations where these two requirements are
met very strongly.      But a closure theory can also be worth pursing even in situations
that are not extremely favorable.    For instance, coarse-grained models of turbulent behavior 
or sub-grid scale parameterizations for complex systems in which  coherent
structures interact with  fluctuations over a range of scales 
may necessitate somewhat crude closures, or perhaps a hierarchy of closures.  
In problems of this kind, our optimization approach furnishes a systematic method
for deriving closures approximations that share the formal structure of thermodynamics.  

As one concrete example we mention an implementation of
our general methodology to the truncated Burgers-Hopf equation  \cite{KT}.
This system has been proposed as a good testbed for model reduction techniques
because it is a relatively simple prototype of turbulent fluid systems \cite{MT1,MT2}.
The continuum dynamics of the Burgers-Hopf PDE is truncated to $n$ Fourier modes, and then
a reduction to the lowest $m \ll n$ modes is considered.     The truncated system, though it has
a noncanonical Hamiltonian structure, possesses a Gaussian invariant measure, and
for $n>20$ its computed
dynamics are observed to be ergodic and mixing and to have nearly Gaussian statistics.
A best-fit closure theory based on Gaussian trial densities produces
explicit, closed reduced equations for the relaxation of the $m$ resolved modes
from nonequilibrium initial conditions. 
These equations have the form of the $m$-mode truncation itself but with 
strong mode-dependent dissipation and modified nonlinear modal interactions. 
Even though this closure has only
one adjustable parameter and the time-scale separation is meager, the predictions
of the best-fit closure with $5$ resolved modes 
agree quite well with direct numerical simulations of large ensembles of solutions
of a $50$-mode dynamics.   Moreover,
this agreement holds for far-from-equilibrium initial conditions as well as near-to-equilibrium
conditions.   
In light of  the structure of the Burgers-Hopf dynamics --- essentially a quadratic nonlinearity
of hydrodynamic type --- these results suggest that more realistic 
systems of this kind may be amenable to a best-fit closure.        

The present paper has focussed exclusively on the relaxation from nonequilibrium
statistical initial conditions of conservative, deterministic dynamical systems.   
It remains to develop the analogous theory for forced systems with nonconservative
or stochastic dynamics.   In the latter context, a promising connection exists between 
our approach and the studies of  nonequilibrium steady states of driven
diffusive systems by Bertini et al.\ \cite{Bertini-etal1,Bertini-etal2,Bertini-etal3}.
In that work  a natural variational principle over paths also arises as well as an 
associated Hamilton-Jacobi equation.
In contrast to our approach, though, these authors are primarily concerned with 
characterizing theoretically the thermodynamic limits of stochastic microscopic models,
not with deriving computationally tractable, predictive reduced models of complex
dynamical systems.     Nonetheless,
 the common features of all these reduction strategies 
and nonequilibrium thermodynamic structures would benefit from further  investigations.

\section{Acknowlegments}

In the course of this work the author benefited from conversations
with R.S.~Ellis, M.~Katsoulakis,  
R.~Kleeman, A.J.~Majda, and P.~Plechac.     
This research was initiated during a sabbatical stay at the
University of Warwick partly supported by an international
short visit fellowship from the Royal Society, and was completed during a 
two-month visit to
the Courant Institute of Mathematical Sciences.   This work received funding from
 the National Science Foundation under grant DMS-0604071.


\vspace{.25cm}

\noi
{\bf Contact Information: }  \\ 

\sf
\noi
\begin{tabular}{ll}
Bruce E. Turkington   \\
Department of Mathematics and Statistics     \\ 
University of Massachusetts   \\
Amherst, MA 01003   \\  
  \\
 \vspace{3mm}   turk@math.umass.edu    
 
\end{tabular}

\end{document}